\documentclass[twocolumn,english,prl,superscriptaddress,longbibliography]{revtex4-1}
\pdfoutput=1
\usepackage[T1]{fontenc}
\usepackage[latin9]{inputenc}
\usepackage{verbatim}
\usepackage{units}
\usepackage{bm}
\usepackage{amsmath}
\usepackage{amssymb}
\usepackage{graphicx}
\usepackage{esint}

\makeatletter
\@ifundefined{definecolor}
 {\usepackage{color}}{}

\@ifundefined{definecolor}{\usepackage{color}}{}
\usepackage{units}\usepackage{bm}

\usepackage{babel}\usepackage{babel}

\usepackage{babel}

\usepackage{babel}

\makeatother

\usepackage{babel}
\begin{document}

\title{Superfluidity in the absence of kinetics in spin-orbit-coupled optical
lattices}

\author{Hoi-Yin Hui}

\affiliation{Department of Physics, Virginia Tech, Blacksburg, Virginia 24061,
USA}

\author{Yongping Zhang}

\affiliation{Department of Physics, Shanghai University, 200444 Shanghai, China}

\author{Chuanwei Zhang}

\affiliation{Department of Physics, The University of Texas at Dallas, Richardson,
Texas 75080, USA}

\author{V. W. Scarola}

\affiliation{Department of Physics, Virginia Tech, Blacksburg, Virginia 24061,
USA}


\date{\today}
\begin{abstract}
At low temperatures bosons typically condense to minimize their single-particle
kinetic energy while interactions stabilize superfluidity. Optical
lattices with artificial spin-orbit coupling challenge this paradigm
because here kinetic energy can be quenched in an extreme regime where
the single-particle band flattens. To probe the fate of superfluidity
in the absence of kinetics we construct and numerically solve interaction-only
tight-binding models in flat bands. We find that novel superfluid
states arise entirely from interactions operating in quenched kinetic
energy bands, thus revealing a distinct and unexpected condensation
mechanism. Our results have important implications for the identification
of quantum condensed phases of ultracold bosons beyond conventional
paradigms. 
\end{abstract}
\maketitle

\section{\noindent Introduction}

\noindent Experiments with ultracold atoms in optical lattices have
opened investigations of strongly correlated systems in extreme regimes,
in particular the extremes of the Fermi-Hubbard and the Bose-Hubbard
models. In the latter case, a quantum phase transition has been observed
\cite{verkerk:1992,jessen:1992,hemmerich:1993}, where the balance
between interaction and single particle kinetics can be tuned to destabilize
the superfluid (SF) into a Mott insulating (MI) state of localized
bosons \cite{jaksch:1998,greiner:2002,bloch:2008}.

Recent theoretical studies have found Bose condensation even in lattice
systems with quenched kinetics that go beyond simply raising optical
lattice depth to decrease inter-site tunneling (hopping). Interesting
consequences of condensation in flat bands include SFs derived solely
from the interaction \cite{huber:2010,You:2012,zhang:2013,Tovmasyan2013}.
However, their experimental implementation faces challenges. For example,
some of the proposals requires long-range hoppings with specific ratios,
while some require spatially varying hopping strengths. These are
difficult to accomplish with ordinary atoms or molecules in simple
optical lattices.

Recently, special optical lattice geometries hosting flat bands have
been proposed and in some cases realized (e.g., hexagonal \cite{soltan-panahi:2011a,tarruell:2012}
and excited bands of kagome lattices \cite{jo:2012,You:2012}). Synthetic
spin-orbit coupling (SOC) \cite{higbie:2002,lin:2011,galitski:2013,Zhai:2015,Wu2015}
has also been found to lead to flat bands \cite{huber:2010,parameswaran:2013,zhang:2013,lin:2014}
on regular lattices with Zeeman fields. It is therefore appropriate
to investigate the characteristics of superfluidity, if it exists
at all, in such flat band systems.

Previous studies of superfluidity in flat bands assumed definite incommensurate
filling ratios which leads to the formation of a condensate. Interactions
are then included at the mean-field level. In realistic experiments
with trapping potentials, it is however more pertinent to ask whether
superfluidity persists for a finite range of chemical potential, or
else the system could phase-separate to regions of MIs with different
particles per site at low hopping. Also, the non-perturbative nature
of flat band systems makes theoretical analyses non-trivial. Understanding
superfluidity in such limits requires unbiased methods, for example
exact diagonalization or density matrix renormalization group (DMRG),
to go beyond mean-field to understand how (or if) superfluidity can
actually occur.

In this paper, we show how to create flat bands using Rashba spin-orbit
coupling in one and two dimensions. We model the band structure and
construct a tight-binding model. With an $s$-wave interaction between
bosons placed in the flat band, we find that even in the absence of
kinetics they condense and form a SF. We compute the mean-field phase
diagram and find competing MI and SF phases (in direct analogy to
what one finds by placing interacting bosons in an ordinary optical
lattice with kinetics). We also go beyond mean-field to probe the
role of quantum fluctuations by using DMRG \cite{white:1992,kuhner:2000,bauer:2011}
to show that the SF survives quantum fluctuations. Our central finding
is that interactions themselves define an effective band structure
in which bosons condense to reveal a new type of SF derived entirely
from interactions that is fundamentally different from SOC SFs that
have been studied up to now (see, e.g., Refs.~\cite{gong:2015,hu:2012,qian:2013,ramachandhran:2012,ramachandhran:2013,sau:2011,wu:2011,zhou:2013}).
The new type of interaction-only SF has distinctive excitations which
can be used to discern it from ordinary SFs.

\begin{figure}
\begin{centering}
\includegraphics[width=1\columnwidth]{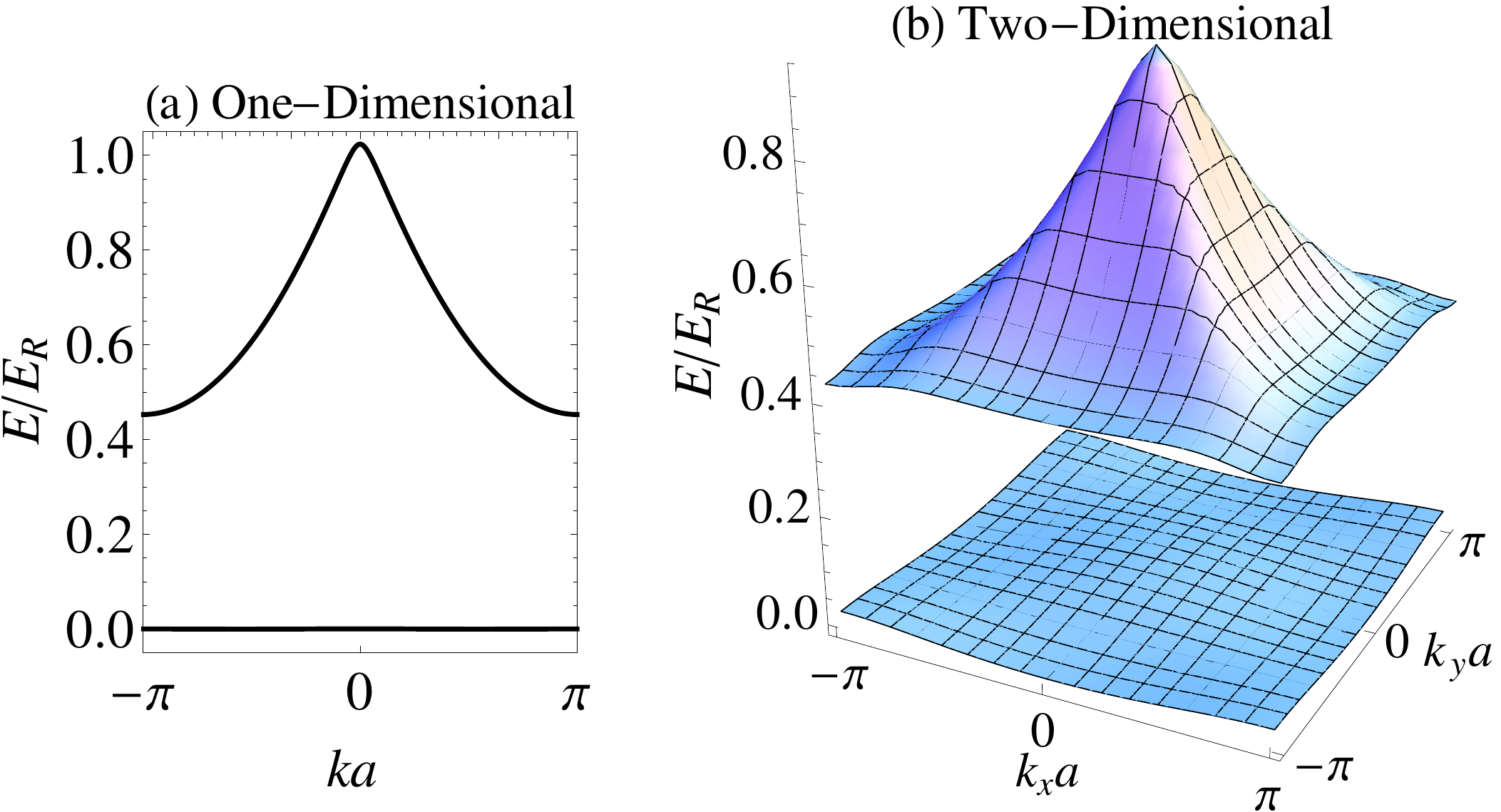} 
\par\end{centering}

\caption{\label{fig:band} Energy versus wave vector for the lowest two energy
bands of Eq.~(\ref{eq:H0}) with $k_{R}=2\pi/a$, $V_{{\rm lat}}=E_{R}$,
for (a) the one-dimensional system, with $\Omega^{*}=8.88E_{R}$,
and (b) the two-dimensional system, with $\Omega^{*}=8.31E_{R}$.}
\end{figure}

\section{\noindent Continuum Model}

\noindent We consider a two-component spin-orbit-coupled Bose-Einstein
condensate in a $d$-dimensional optical lattice, described by the
Hamiltonian 
\begin{eqnarray}
\hat{H} & = & \int d\bm{r}\hat{b}^{\dagger}\left(\bm{r}\right)H_{0}\left(\bm{r}\right)\hat{b}\left(\bm{r}\right)\nonumber \\
 &  & +\frac{U_{0}}{2}\int d\bm{r}\sum_{\sigma\sigma'}\hat{b}_{\sigma}^{\dagger}\left(\bm{r}\right)\hat{b}_{\sigma'}^{\dagger}\left(\bm{r}\right)\hat{b}_{\sigma'}\left(\bm{r}\right)\hat{b}_{\sigma}\left(\bm{r}\right)\label{eq:H}\\
H_{0} & = & \frac{\hbar^{2}\bm{k}^{2}}{2m}+\frac{\hbar k_{R}}{m}\bm{{\cal F}}\cdot\bm{\sigma}+\Omega\sigma_{z}\nonumber \\
 &  & +V_{{\rm lat}}\sum_{i=1}^{d}\sin^{2}\frac{\pi\bm{r}\cdot\bm{e}_{i}}{a},\label{eq:H0}
\end{eqnarray}
 with $\hat{b}^{\dagger}=\left(\hat{b}_{\uparrow}^{\dagger},\hat{b}_{\downarrow}^{\dagger}\right)$
where $\hat{b}_{\sigma}^{\dagger}\left(\bm{r}\right)$ creates a particle
with spin $\sigma\in\left\{ \uparrow,\downarrow\right\} $ at position
$\bm{r}$ (with unit vectors $\bm{e}_{i}$ defining chain and square
lattices for $d=1$ and 2, respectively), and $U_{0}$ is the $s$-wave
interaction strength. In the single-particle Hamiltonian, $H_{0}$,
$m$ is the mass of each particle, $\bm{k}$ is the momentum operator,
$k_{R}$ characterizes the strength of the SOC induced by the Raman
lasers, $\Omega$ is the Rabi frequency which acts as the Zeeman field,
and $V_{{\rm lat}}$ is the depth of the optical lattice. In one dimension:
$\bm{k}=\bm{{\cal F}=}-i\partial_{x}$ and $\bm{\sigma}=\sigma_{x}$,
while in the two dimensions: $\bm{k}=\left(-i\partial_{x},-i\partial_{y}\right)$,
$\bm{{\cal F}}=\left(i\partial_{y},-i\partial_{x}\right)$ and $\bm{\sigma}=\left(\sigma_{x},\sigma_{y}\right)$,
in which the Pauli matrices $\bm{\sigma}$ act on the spin sectors
of $b^{\dagger}$. For convenience, we define the lattice recoil energy
$E_{R}\equiv(\pi/a)^{2}\hbar^{2}/(2m)$ to express some of the parameters.

\section{\noindent Tight-binding Model}

\begin{figure}
\begin{centering}
\includegraphics[width=1\columnwidth]{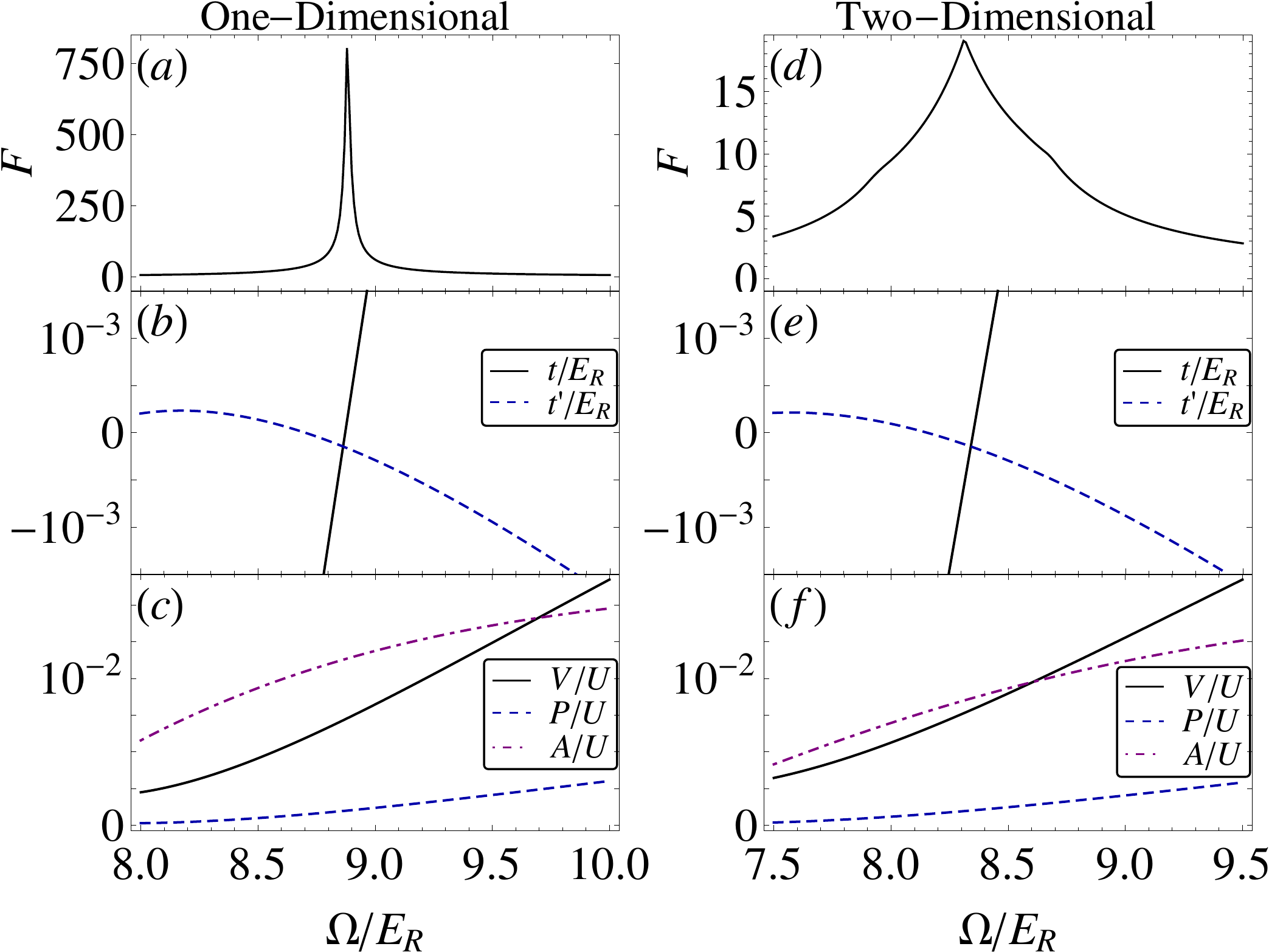} 
\par\end{centering}

\caption{\label{fig:TB}(a) The flatness ratio (defined as the ratio of the
gap between the two lowest bands to the width of the lowest band \cite{zhang:2013})
versus Rabi frequency for $k_{R}=2\pi/a$ and $V_{{\rm lat}}=E_{R}$
in one dimension. (b) The hopping parameters $t$ and $t'$ against
the Rabbi frequency. (c) The ratio of interaction parameters $V$,
$P$ and $A$ to $U$. The right column (d-f) plots the same quantities
for a two-dimensional system (with the same $k_{R}$ and $V_{{\rm lat}}$).
Choosing $\Omega$ to lie at the peak leads to Eq.~\eqref{eq:HTBreduced}.}
\end{figure}

\noindent We reduce the above continuum model to a tight-binding model
and project the interactions to the flat Bloch band. In the absence
of interaction, the Bloch functions $\psi_{\bm{k}}\left(\bm{r}\right)=u_{\bm{k}}\left(\bm{r}\right)e^{i\bm{k}\cdot\bm{r}}$
are found by expanding $u_{\bm{k}}\left(\bm{r}\right)$ in plane waves
with periodicity commensurate with that of the lattice. For given
$k_{R}$ and $V_{{\rm lat}}$, an optimal value of $\Omega$, $\Omega^{*}$,
produces a lowest band with the highest flatness ratio \cite{zhang:2013}.
Fig.~\ref{fig:band} shows the band structures in $d=1$ and $2$
where $\Omega=\Omega^{*}$, with $k_{R}=2\pi/a$ and $V_{{\rm lat}}=E_{R}$.
The dependence of $F$ on $\Omega$ for the same parameters is plotted
in Figs.~\ref{fig:TB}(a) and (d), which shows that a high flatness
ratio {[}the large peaks in Figs.~\ref{fig:TB}(a) and (d){]} is
achievable with moderate parameter strengths.

We construct the Wannier functions to obtain the tight-binding model.
We define a two-component Wannier function localized at cell $\bm{R}_{i}$,
$w\left(\bm{r}-\bm{R}_{i}\right)=\left[w_{\uparrow}\left(\bm{r}-\bm{R}_{i}\right),w_{\downarrow}\left(\bm{r}-\bm{R}_{i}\right)\right]^{T}$,
with $w_{\sigma}\left(\bm{r}-\bm{R}_{i}\right)=\sum_{\bm{k}}e^{i\bm{k}\cdot\left(\bm{r}-\bm{R}_{i}\right)}u_{\sigma\bm{k}}\left(\bm{r}\right)$,
where $u_{\sigma\bm{k}}\left(\bm{r}\right)$ are the Bloch functions
for the lowest band. The phases of the Bloch functions are fixed by
requiring the spread of the Wannier function, $\left\langle \bm{r}^{2}\right\rangle -\left\langle \bm{r}\right\rangle ^{2}$
(where $\left\langle \bm{r}^{l}\right\rangle \equiv\sum_{\sigma\bm{k}}\left\langle u_{\sigma\bm{k}}\right|\left(i\nabla_{\bm{k}}\right)^{l}\left|u_{\sigma\bm{k}}\right\rangle $),
to be minimized \cite{marzari:1997}. The tight-binding model is constructed
by effecting the transformation to the flat-band spinor basis states:
$\hat{a}_{i}^{\dagger}=\Sigma_{\sigma}\int d\bm{r}w_{\sigma}\left(\bm{r}-\bm{R}_{i}\right)\hat{b}_{\sigma}^{\dagger}\left(\bm{r}\right)$
onto Eq.~\eqref{eq:H}, with which the tight-binding parameters can
then be readily computed by taking the overlaps of $w\left(\bm{r}\right)$
(see App.~\ref{sec:TBparameters}).

The non-interacting part of the Hamiltonian leads to hopping terms
($-t\sum_{\left\langle ij\right\rangle }\hat{a}_{i}^{\dagger}\hat{a}_{j}+{\rm h.c.}$,
where $\left\langle ij\right\rangle $ denotes nearest-neighbors,
and $-t'\sum_{\left\langle \left\langle ij\right\rangle \right\rangle }\hat{a}_{i}^{\dagger}\hat{a}_{j}+{\rm h.c.}$,
where $\left\langle \left\langle ij\right\rangle \right\rangle $
denotes next-nearest-neighbors) and the chemical potential term ($-\mu\sum_{i}\hat{a}_{i}^{\dagger}\hat{a}_{i}$).
For a range of parameter values, we have numerically computed $t$,
$t'$ and $\mu$ and verified that the band dispersion resulting from
these terms agrees very well with the band structure obtained directly
from the plane-wave expansion of Eq.~\eqref{eq:H0} (to within 5\%),
indicating the adequacy of our tight-binding approximation. At $k_{R}=2\pi/a$
and $V_{{\rm lat}}=E_{R}$, the values of $t$ and $t'$ are plotted
against $\Omega$ in Figs.~\ref{fig:TB}(b) and \ref{fig:TB}(e).
At the optimal flatness point $\left(\Omega=\Omega^{*}\right)$, $t$
and $t'$ are vanishingly small ($<10^{-4}E_{R}$, compared with $t=0.178E_{R}$
at $\Omega=0$), and are much smaller than the density-assisted hopping
$A$ (for example, for $^{87}{\rm Rb}$ atoms with $k_{R}=2\pi/a$
in a lattice with $a=384{\rm nm}$ and $V_{{\rm lat}}=E_{R}$ \cite{Wu2015},
$t'/A\sim t/A\sim0.075$ at the optimal point, where the density-assisted
term proportional to $A$ is defined in the next paragraph). This
motivates us to drop the hopping terms in the effective tight-binding
model and investigate the resulting interacting flat band model.

When truncated to the nearest-neighbor terms, the interaction {[}$U_{0}$
in Eq.~(\ref{eq:H}){]} in general leads to four terms in the tight-binding
model, which are the on-site interaction $(U/2)\sum_{i}\hat{n}_{i}\left(\hat{n}_{i}-1\right)$
(where $\hat{n}_{i}=\hat{a}_{i}^{\dagger}\hat{a}_{i}$), nearest-neighbor
interaction $V\sum_{\left\langle ij\right\rangle }\hat{n}_{i}\hat{n}_{j}$,
density-assisted hopping $-A\sum_{\left\langle ij\right\rangle }\left[\hat{a}_{j}^{\dagger}\left(\hat{n}_{i}+\hat{n}_{j}\right)\hat{a}_{i}+{\rm h.c.}\right]$
and pair hopping $P\sum_{\left\langle ij\right\rangle }\left(\hat{a}_{i}^{\dagger}\hat{a}_{i}^{\dagger}\hat{a}_{j}\hat{a}_{j}+{\rm h.c.}\right)$.
Their dependencies on $\Omega$ for $k_{R}=2\pi/a$ and $V_{{\rm lat}}=E_{R}$
are plotted in Figs.~\ref{fig:TB}(c) and \ref{fig:TB}(f). It is
important to stress that in a ``trivially'' flat band where hoppings
($t$ and $t'$) are suppressed by increasing the depth of the lattice,
both $A$ and $V$ are also suppressed, making the system classical.
In contrast, for the system we are investigating, only the hoppings
are suppressed.

Since $P$ is much smaller than $V$ or $A$ near $\Omega^{*}$, we
drop the pair-hopping term in the tight-binding model. We also note
that near $\Omega^{*}$, $V$ and $A$ have similar values. This leads
us to study the interaction-only tight-binding model: 
\begin{eqnarray}
\hat{H}_{TB} & = & -\mu\sum_{i}\hat{n}_{i}+\frac{U}{2}\sum_{i}\hat{n}_{i}\left(\hat{n}_{i}-1\right)+V\sum_{\left\langle ij\right\rangle }\hat{n}_{i}\hat{n}_{j}\nonumber \\
 &  & -A\sum_{\left\langle ij\right\rangle }\left[\hat{a}_{j}^{\dagger}\left(\hat{n}_{i}+\hat{n}_{j}\right)\hat{a}_{i}^{\vphantom{\dagger}}+{\rm h.c.}\right],\label{eq:HTBreduced}
\end{eqnarray}
where $V\approx A\approx0.01U$ for the parameters we have chosen
(see Fig.~\ref{eq:HTBreduced}). $V$ and $A$ can be varied with
$V_{{\rm lat}}$ and $k_{R}$, and the corresponding optimal value
of $\Omega$, in Eq.~\eqref{eq:H0}. For simplicity and in line with
what we observe in Fig.~\ref{fig:TB}, we set $A=V$ in the rest
of our numerical study. We have checked that slight deviations from
this condition do not qualitatively alter the phase diagrams presented.

\begin{figure}
\begin{centering}
\includegraphics[width=0.75\columnwidth]{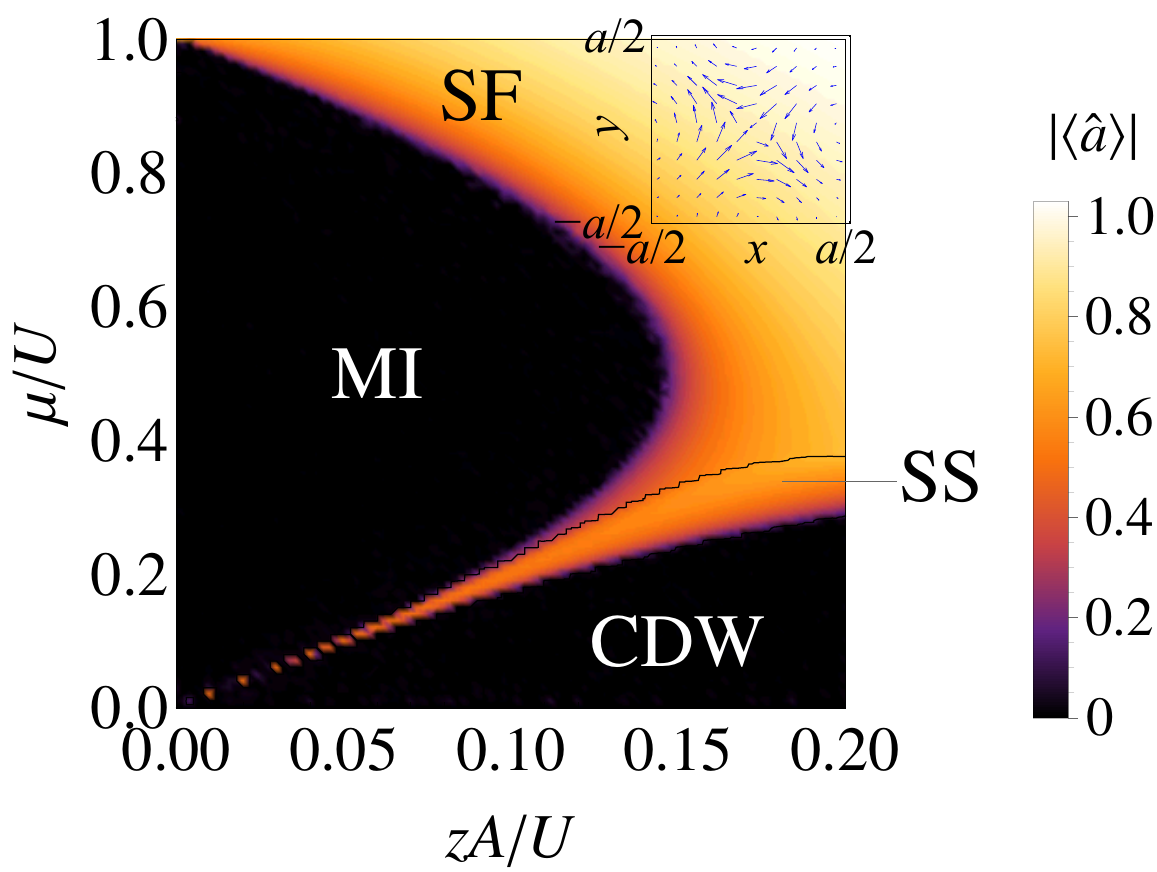} 
\par\end{centering}

\caption{\label{fig:MF}The magnitude of the SF order parameter, $\left|\left\langle \hat{a}\right\rangle \right|$,
against $\mu$ and $A(=V)$ in the model Eq.~(\ref{eq:HTBreduced}).
$z=2$ $(z=4)$ is the coordination number in one (two) dimensions.
The MI and CDW phases have $\left|\left\langle \hat{a}\right\rangle \right|=0$
and the SF and SS phases have $\left|\left\langle \hat{a}\right\rangle \right|\neq0$.
The boundary between the SF and the SS phase is set at ${\rm var}\left\langle \hat{a}\right\rangle =0.001$.
The inset shows the superfluid spin texture (App.~\ref{sec:SpinTextures})
in a unit cell for the two-dimensional system in the superfluid phase
at its optimal flatness point as in Fig.~\ref{fig:band}b, leading
to $V\approx A\approx0.01U$. }
\end{figure}

\section{\noindent Mean-field Phase Diagram}

\noindent We now turn to an analysis of the phases of Eq.~\eqref{eq:HTBreduced}.
We first adopt a mean-field approach which ignores quantum fluctuations.
Quantum fluctuations become more important in low-dimensions. In the
following section we shall examine the role of quantum fluctuations
using DMRG in one dimension. We will show that the mean-field approach
presented in this section gives qualitatively correct results.

We construct the mean-field phase diagram of Eq.~\eqref{eq:HTBreduced}
using the Gutzwiller ansatz wavefunction, $\left|\Psi\right\rangle =\prod_{i}\sum_{n}f_{n}^{i}\left|n\right\rangle _{i}$
(where $\left|n\right\rangle _{i}$ is the Fock state with $n$ bosons
at the $i^{{\rm th}}$ site) \cite{rokhsar:1991,jaksch:1998}. We
obtain the mean-field ground state by minimizing $\left\langle \Psi\right|\hat{H}_{TB}\left|\Psi\right\rangle $
with respect to $f_{n}^{i}$. To characterize the ground state, we
compute the average SF order parameter$\left\langle \hat{a}\right\rangle $,
its spatial variance ${\rm var}\left\langle \hat{a}\right\rangle $,
and the spatial variance of the occupation number ${\rm var}\left\langle \hat{n}\right\rangle $.
With these we can identify the MI phase (with $\left\langle \hat{a}\right\rangle =0$
and ${\rm var}\left\langle \hat{n}\right\rangle =0$), the charge
density wave (CDW) phase (with $\left\langle \hat{a}\right\rangle =0$
and ${\rm var}\left\langle \hat{n}\right\rangle \neq0$), the supersolid
(SS) phase (with $\left\langle \hat{a}\right\rangle \neq0$ and ${\rm var}\left\langle \hat{a}\right\rangle \neq0$),
and the SF phase (with $\left\langle \hat{a}\right\rangle \neq0$
and ${\rm var}\left\langle \hat{a}\right\rangle =0$).

The resulting phase diagram is shown in Fig.~\ref{fig:MF}, with
MI, CDW and SS phases at small $A$ and SF at large $A$. Its resemblance
to that of the conventional extended Bose-Hubbard model \cite{fisher:1997,batrouni:1995,vanOtterlo:1995,kuhner:2000}
can be understood in a mean-field decoupling of the density-assisted
hopping term {[}proportional to $A$ in Eq.~\eqref{eq:HTBreduced}{]}:
\[
\hat{a}_{i}^{\dagger}\hat{n}_{i}\hat{a}_{j}^{\vphantom{\dagger}}\rightarrow\left\langle \hat{n}\right\rangle \hat{a}_{i}^{\dagger}\hat{a}_{j}.
\]
 Here the density-assisted hopping plays the role of conventional
hopping to yield an effective band structure (with an effective hopping
of strength $\langle\hat{n}\rangle A$). The mean-field phase diagram
indicates that bosons still condense and form a superfluid phase even
in the absence of kinetics. After condensing into the band minimum
of the effective band, the residual interactions support the formation
of a superfluid.

There are similarities and differences between the superfluid discussed
here and the superfluids typically discussed in the ordinary Bose-Hubbard
model of optical lattices with SOC. The mean-field superfluid order
parameter $\left\langle \hat{a}\right\rangle $ defines a spinor when
decomposed in terms of the original spinful bosons since $\langle\hat{a}_{i}\rangle=\int d\bm{r}[w_{\uparrow}\left(\bm{r}-\bm{R}_{i}\right)\langle\hat{b}_{\uparrow}\left(\bm{r}\right)\rangle+w_{\downarrow}\left(\bm{r}-\bm{R}_{i}\right)\langle\hat{b}_{\downarrow}\left(\bm{r}\right)\rangle]$.
The superfluid discussed here therefore has a canted spin structure
which is to be expected from SOC coupling. But there are fundamental
differences. The strength of superfluidity is determined almost entirely
from interactions (not a competition between kinetics and interactions).
Furthermore, the condensation of bosons occurs in an effective band
with a density-dependent strength. But most importantly, the excitations
of the superfluid discussed here are multi-particle because they derive
entirely from interactions. The observable consequences of the differences
will be discussed in the summary. We now address the role of quantum
fluctuations.

\begin{figure}
\begin{centering}
\includegraphics[width=1\columnwidth]{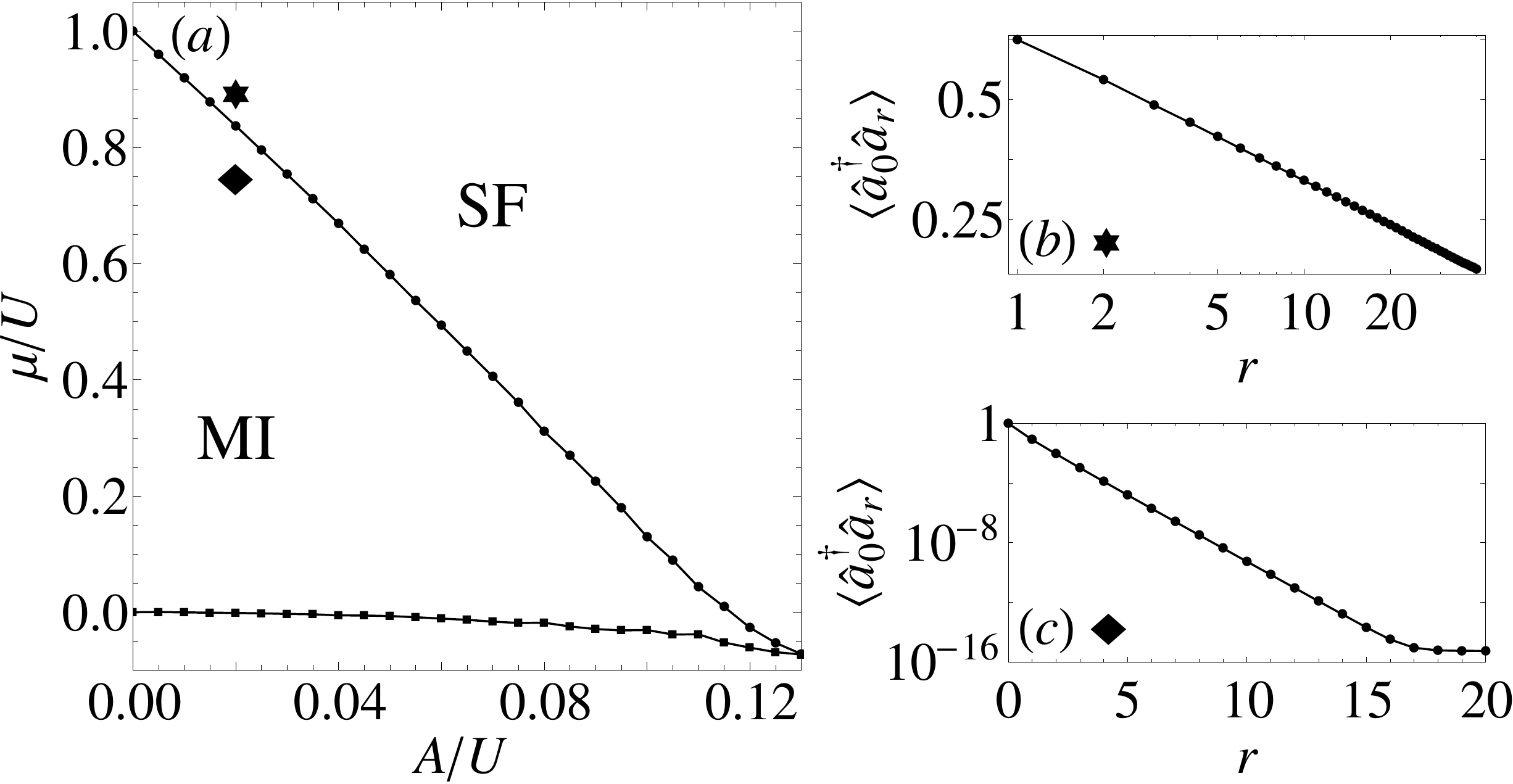} 
\par\end{centering}

\caption{\label{fig:DMRG}(a) The gapped phase (MI with 1 particle per site)
boundary defined by plotting the chemical potential against $A(=V)$
for Eq.~\eqref{eq:HTBreduced}. (b) The single particle density matrix,
$\langle\hat{a}_{0}^{\dagger}\hat{a}_{r}\rangle$, versus distance
$r$ in the compressible phase {[}star in (a), with $\mu=U$ and $A=V=0.01U${]}
on a log-log scale. (c) The same as (b) but in the gapped phase {[}diamond
in (a), with $\mu=U$ and $A=V=0.01U${]} in the log-linear scale.
The lines are a guide to the eye. Note that the chosen values of $A$
in (b) and (c) are consistent with the estimated values for ${\rm Rb}^{87}$.}
\end{figure}

\section{\noindent Phases in Low Dimensions}

\noindent To address the impact of quantum fluctuations on our phase
diagram, we pass to a regime where quantum fluctuations are strongest:
one dimension. Here we can use the DMRG method to compute what is
essentially the exact phase diagram \cite{kuhner:2000} so as to complement
our mean-field results above. We will show qualitative agreement between
our mean-field results for higher dimensions and our one-dimensional
DMRG results. This shows that quantum fluctuations do not qualitatively
change the conclusions drawn above.

To find the phase boundaries, we compute the ground state energy of
the model on a chain of length $L$ with integer filling, $E_{L}^{(0)}$,
and compare this with the ground state energies with $L\pm1$ particles,
$E_{L}^{\left(\pm\right)}$. The upper and lower chemical potential
boundaries ($\mu_{\pm}$) of the gapped phase are $\mu_{\nicefrac{+}{-}}=\lim_{L\rightarrow\infty}\left[E_{L}^{\left(\nicefrac{+}{0}\right)}-E_{L}^{\left(\nicefrac{0}{-}\right)}\right]$
where the limits are numerically computed by extrapolation from finite-$L$
results. The upper boundary for the $\nu=1$ Mott lobe obtained from
this approach are plotted in Fig.~\ref{fig:DMRG}(a). Here we see
MI and superfluid phases expected from the mean-field theory. Quantum
fluctuations force the otherwise rounded MI lobe to instead converge
to a Berezinskii-Kosterlitz-Thouless point at the tip of the lobe
\cite{kuhner:1998}.

For an unambiguous identification of the gapped and the compressible
phases, we also compute the off-diagonal order (ODO), $\langle\hat{a}_{0}^{\dagger}\hat{a}_{r}^{\vphantom{\dagger}}\rangle$,
for representative points in the two phases. Fig.~\ref{fig:DMRG}(b)
plots the ODO against distance $r$ for the compressible phase in
a log-log scale. Since the data falls roughly on a straight line,
it shows that the ODO decays algebraically, which is indicative of
a superfluid phase in one dimension. In contrast, the ODO decays exponentially
in the gapped phase {[}log-linear plot of Fig.~\ref{fig:DMRG}(c){]}.
Superfluidity is absent there and the phase is a MI.

Our results show qualitative agreement between the mean-field phase
diagram {[}Fig.~\ref{fig:MF}{]} and the DMRG results in {[}Fig.~\ref{fig:DMRG}{]}.
The presence of a superfluid phase in both phase diagrams shows that
quantum fluctuations should preserve the MI and superfluid even in
two dimensions since quantum fluctuations are less severe in two dimensions.

\section{\noindent Discussion}

\noindent The parameter regime discussed here is accessible with current
setups. For example, a recent experiment \cite{hamner:2015} with
$^{87}$Rb implemented spin-orbit coupling strengths and optical lattice
depths ($k_{R}=1.96\pi/a$ and $V_{{\rm lat}}=1.4E_{R}$, respectively)
in one dimension, near the regime considered here ($k_{R}=2\pi/a$
and $V_{{\rm lat}}=1E_{R}$). Another experiment \cite{Wu2015} implemented
SOC in two-dimensions with $^{87}{\rm Rb}$ atoms. Using the experimental
values of $a=384{\rm nm}$ and assuming the perpendicular confinement
of strength $V_{{\rm lat},\bot}=81E_{R}$, we find $t\approx2{\rm Hz}$,
$A\approx27{\rm Hz}$, and $U\approx3{\rm kHz}$. These realistic
parameters lead to a ratio $A\approx V\approx0.01U$, which is strong
enough to reveal the MI-superfluid transition. Fig.~\ref{fig:DMRG}(b)
and (c) demonstrate the superfluid and Mott phases realized under
these values..

The novel superfluid phase discussed here has unique experimental
identifiers that contrast from the conventional superfluid states
observed in optical lattice experiments \cite{bloch:2008}. We first
note that here the superfluid-MI transition occurs at a very low lattice
depth, $V_{{\rm lat}}\sim1E_{R}$. Furthermore, the superfluid-MI
phase boundaries are defined by the interaction strength which depends
on the lattice depth only insofar as the Wannier functions modify
the $s$-wave scattering contribution to $U$ and $A$. Second, we
point out that since the physics discussed here is driven by density-assisted
tunneling, methods to observe density-assisted tunneling \cite{baier:2016}
could be implemented to prove the dominance of this process. But the
most important distinction stems from the unique nature of the interaction-only
superfluid itself. Here the excitations must occur in the two-particle
sector since they derive entirely from two-particle interactions.
As such, probes of excitations should have unique signatures. For
example, the momentum distribution peaks in superfluid phases \cite{bloch:2008}
show additional structure due to particle-hole excitations -- the
visibility (defined as $\frac{n_{{\rm max}}-n_{{\rm min}}}{n_{{\rm max}}+n_{{\rm min}}}$,
where $n_{{\rm max}/{\rm min}}$ are the maximum/minimum intensity
along the circle $\left|k\right|=k_{R}$ in the momentum distribution
\cite{gerbier:2005,gerbier:2005a}), scales linearly with the density
$\approx(\langle\hat{n}\rangle+1)4zt/3U$, where $z$ is the coordination
for ordinary superfluid \cite{gerbier:2005,gerbier:2005a}. On the
other hand, since the superfluid discussed here is derived solely
from the interaction ($A\hat{a}_{j}^{\dagger}\hat{n}_{i}\hat{a}_{i}^{\vphantom{\dagger}}$),
its visibility should scale with \emph{square} of the density $\approx(\langle\hat{n}\rangle+1)\langle\hat{n}\rangle4zA/3U$.
(The position of the peaks should be the same as that of ordinary
superfluid provided that the density is uniform.) Other more local
probes are also possible. The realistic parameters we have ($t/U<0.001$)
would normally result in a Mott state at the trap center if the assisted
tunneling term ($A$) were absent. But the observation of a finite
condensate fraction near the trap center would yield strong evidence
for the superfluid discussed here. Another, more direct example, would
employ atomic gas microscopes. These setups \cite{bakr:2009a,weitenberg:2011,miranda:2015,cheuk:2012}
offer direct probes of the dynamics under the effective Hamiltonian
and would reveal the unique two-particle nature of the excitations.

It is widely understood that bosons condense into the lowest single-particle
kinetic energy state while interactions perturb BECs into a superfluid
state. We have studied optical lattice bosons in a flat spin-orbit
band generated by Rashba spin-orbit coupling. We have derived and
solved an interaction-only tight-binding model to show that even in
the absence of kinetics the interaction itself leads to an effective
band that allows condensation and the formation of a superfluid. 
\begin{acknowledgments}
VWS and HH acknowledge support from AFOSR (FA9550-15-1-0445) and ARO
(W911NF-16-1-0182). CZ is supported by ARO (W911NF-12-1-0334) and
NSF (PHY-1505496).
\end{acknowledgments}
\appendix

\section{Maximally-Localized Wannier Functions}

We briefly review the computational procedure for obtaining the maximally
localized wavefunction for the lowest band. We follow closely the
treatment of Ref.~\cite{marzari:1997} for isolated bands on a square
lattice.

By using Bloch's theorem, the single-particle wavefunction in the
lattice can be written as $\psi_{\sigma\bm{k}}\left(\bm{r}\right)=\sum_{\bm{K}}c_{\sigma,\bm{k}-\bm{K}}e^{i\left(\bm{k}-\bm{K}\right)\cdot\bm{r}}$,
where $\bm{k}$ is the crystal momentum and $\bm{K}$ are the reciprocal
lattice vectors. Substituting this form of $\psi_{\sigma\bm{k}}\left(\bm{r}\right)$
into $H_{0}\psi={\cal E}\psi$ yields

\begin{eqnarray}
 & \big[ & \frac{\hbar^{2}}{2m}\left(\bm{k}-\bm{K}\right)^{2}+\frac{\hbar k_{R}}{m}\left(\left(k_{x}-K_{x}\right)\sigma_{y}-\left(k_{y}-K_{y}\right)\sigma_{x}\right)\nonumber \\
 & + & \Omega\sigma_{z}\big]c_{\bm{k}-\bm{K}}+\sum_{\bm{K}'}U_{\bm{K}'-\bm{K}}c_{\bm{k}-\bm{K}'}={\cal E}c_{\bm{k}-\bm{K}},
\end{eqnarray}
 where $c_{\bm{k}-\bm{K}}=\left(c_{\uparrow\bm{k}-\bm{K}},c_{\downarrow\bm{k}-\bm{K}}\right)^{T}$,
and 
\begin{equation}
U_{\bm{K}}=\frac{V_{{\rm lat}}}{a^{2}}\int_{0}^{a}\int_{0}^{a}dxdy\left(\sin^{2}\frac{\pi x}{a}+\sin^{2}\frac{\pi y}{a}\right)e^{-i\left(K_{x}x-K_{y}y\right)}.
\end{equation}

For each value of $\bm{k}$, we solve the ground state to construct
the wavefunction. To fix the phases of the Bloch wavefunctions, we
now rewrite the wavefunction as $\psi_{\sigma\bm{k}}\left(\bm{r}\right)=e^{i\phi\left(\bm{k}\right)}u_{\sigma\bm{k}}\left(\bm{r}\right)e^{i\bm{k}\cdot\bm{r}}$,
where $u_{\sigma\bm{k}}\left(\bm{r}\right)=\sum_{\bm{K}}c_{\sigma,\bm{k}-\bm{K}}e^{-i\bm{K}\cdot\bm{r}}$
and the phases $\phi\left(\bm{k}\right)$ are to be fixed. Following
Ref.~\cite{marzari:1997} we define 
\begin{align}
\tilde{M}^{\left(\bm{k},\bm{b}\right)} & =e^{-i\phi\left(\bm{k}\right)+i\phi\left(\bm{k}+\bm{b}\right)}\left\langle u_{\bm{k}}|u_{\bm{k}+\bm{b}}\right\rangle \\
 & \equiv e^{-i\phi\left(\bm{k}\right)+i\phi\left(\bm{k}+\bm{b}\right)}M^{\left(\bm{k},\bm{b}\right)}
\end{align}
 where the inner product is defined as the summation over the spin
index followed by an integral over a unit cell, and $u_{\bm{k}+\bm{b}}$
is normalized such that $\left\langle u_{\bm{k}}|u_{\bm{k}+\bm{b}}\right\rangle =1$.
The vectors $\bm{b}$ points to the four nearest-neighbors (or two
in one dimension) in the discretization of the Brillouin zone.

The phases are fixed by requiring $G^{\left(\bm{k}\right)}=\sum_{\bm{b}}{\rm Im}{\rm ln}\tilde{M}^{\left(\bm{k},\bm{b}\right)}$
to vanish identically for all $\bm{k}$ \cite{marzari:1997}. Written
in terms of $\phi\left(\bm{k}\right)$ and $M^{\left(\bm{k},\bm{b}\right)}$,
we have 
\begin{equation}
\sum_{\bm{b}}\left[\phi\left(\bm{k}+\bm{b}\right)-\phi\left(\bm{k}\right)\right]=\sum_{\bm{b}}{\rm Im}{\rm ln}M^{\left(\bm{k},\bm{b}\right)},
\end{equation}
 which is recognized as a (discretized) Poisson equation for the phase
field $\phi\left(\bm{k}\right)$, and is amenable to standard numerical
treatments.

\section{Tight-binding parameters\label{sec:TBparameters}}

By substituting the transformation 
\begin{equation}
\hat{b}_{\sigma}\left(\bm{r}\right)=\sum_{i}w_{\sigma}\left(\bm{r}-\bm{R}_{i}\right)\hat{a}_{i}\label{eq:atob}
\end{equation}
 into $\hat{H}$ and ignoring integrals involving Wannier functions
with more than two lattice sites apart, the Hamiltonian becomes

\begin{eqnarray}
\hat{H} & \approx & -t\sum_{\left\langle ij\right\rangle }\hat{a}_{i}^{\dagger}\hat{a}_{j}-t'\sum_{\left\langle \left\langle ij\right\rangle \right\rangle }\hat{a}_{i}^{\dagger}\hat{a}_{j}+{\rm h.c.}-\mu\sum_{i}\hat{a}_{i}^{\dagger}\hat{a}_{i}\nonumber \\
 &  & +\frac{U}{2}\sum_{i}\hat{n}_{i}\left(\hat{n}_{i}-1\right)+V\sum_{\left\langle ij\right\rangle }\hat{n}_{i}\hat{n}_{j}\nonumber \\
 &  & -A\sum_{\left\langle ij\right\rangle }\left[\hat{a}_{j}^{\dagger}\left(\hat{n}_{i}+\hat{n}_{j}\right)\hat{a}_{i}+{\rm h.c.}\right]\nonumber \\
 &  & +P\sum_{\left\langle ij\right\rangle }\left(\hat{a}_{i}^{\dagger}\hat{a}_{i}^{\dagger}\hat{a}_{j}\hat{a}_{j}+{\rm h.c.}\right)
\end{eqnarray}
 with tight-binding parameters: 
\begin{eqnarray*}
\mu & = & -\int d\bm{r}w^{\dagger}\left(\bm{r}\right)H_{0}w\left(\bm{r}\right)\\
t & = & -\int d\bm{r}w^{\dagger}\left(\bm{r}\right)H_{0}w\left(\bm{r}-\bm{r}_{1}\right)\\
t' & = & -\int d\bm{r}w^{\dagger}\left(\bm{r}\right)H_{0}w\left(\bm{r}-\bm{r}_{2}\right)\\
U & = & U_{0}\int d\bm{r}\sum_{\sigma\sigma'}w_{\sigma}^{*}\left(\bm{r}\right)w_{\sigma'}^{*}\left(\bm{r}\right)w_{\sigma'}\left(\bm{r}\right)w_{\sigma}\left(\bm{r}\right)\\
V & = & 2U_{0}\int d\bm{r}\sum_{\sigma\sigma'}w_{\sigma}^{*}\left(\bm{r}\right)w_{\sigma'}^{*}\left(\bm{r}-\bm{r}_{1}\right)w_{\sigma'}\left(\bm{r}-\bm{r}_{1}\right)w_{\sigma}\left(\bm{r}\right)\\
A & = & -U_{0}\int d\bm{r}\sum_{\sigma\sigma'}w_{\sigma}^{*}\left(\bm{r}\right)w_{\sigma'}^{*}\left(\bm{r}\right)w_{\sigma'}\left(\bm{r}\right)w_{\sigma}\left(\bm{r}-\bm{r}_{1}\right)\\
P & = & \frac{U_{0}}{2}\int d\bm{r}\sum_{\sigma\sigma'}w_{\sigma}^{*}\left(\bm{r}\right)w_{\sigma'}^{*}\left(\bm{r}\right)w_{\sigma'}\left(\bm{r}-\bm{r}_{1}\right)w_{\sigma}\left(\bm{r}-\bm{r}_{1}\right)
\end{eqnarray*}
 where $\bm{r}_{1}=a\hat{e}_{x}$ and $\bm{r}_{2}=\begin{cases}
2a\hat{e}_{x}, & d=1\\
a\hat{e}_{x}+a\hat{e}_{y}, & d=2
\end{cases}$. By computing the maximally localized Wannier functions we are therefore
able to find all tight-binding parameters.

\section{Spin Textures\label{sec:SpinTextures}}

To compute the spin texture $\langle\hat{b}^{\dagger}\vec{\sigma}\hat{b}\rangle$
for the ground state, we use Eq.~\eqref{eq:atob} to express the
$\hat{b}$ operators in terms of the $\hat{a}$ operators. We then
apply the mean field decoupling for both the Mott and superfluid phases.
In the Mott state we have: 
\begin{align}
\left\langle \hat{b}_{\alpha}^{\dagger}\left(\bm{r}\right)\hat{b}_{\beta}\left(\bm{r}\right)\right\rangle  & =\sum_{i,j}w_{\alpha}^{*}\left(\bm{r}-\bm{R}_{i}\right)w_{\beta}\left(\bm{r}-\bm{R}_{j}\right)\left\langle \hat{a}_{i}^{\dagger}\hat{a}_{j}\right\rangle \nonumber \\
 & =\sum_{i,j}w_{\alpha}^{*}\left(\bm{r}-\bm{R}_{i}\right)w_{\beta}\left(\bm{r}-\bm{R}_{j}\right)\delta_{ij}\nonumber \\
 & =\sum_{i}w_{\alpha}^{*}\left(\bm{r}-\bm{R}_{i}\right)w_{\beta}\left(\bm{r}-\bm{R}_{i}\right).
\end{align}
 On the other hand, in the superfluid phase, we have: 
\[
\left\langle \hat{b}_{\alpha}^{\dagger}\left(\bm{r}\right)\hat{b}_{\beta}\left(\bm{r}\right)\right\rangle =\sum_{i,j}w_{\alpha}^{*}\left(\bm{r}-\bm{R}_{i}\right)w_{\beta}\left(\bm{r}-\bm{R}_{j}\right)\left|\left\langle \hat{a}\right\rangle \right|^{2}.
\]

\begin{figure}
\begin{centering}
\includegraphics[width=0.75\columnwidth]{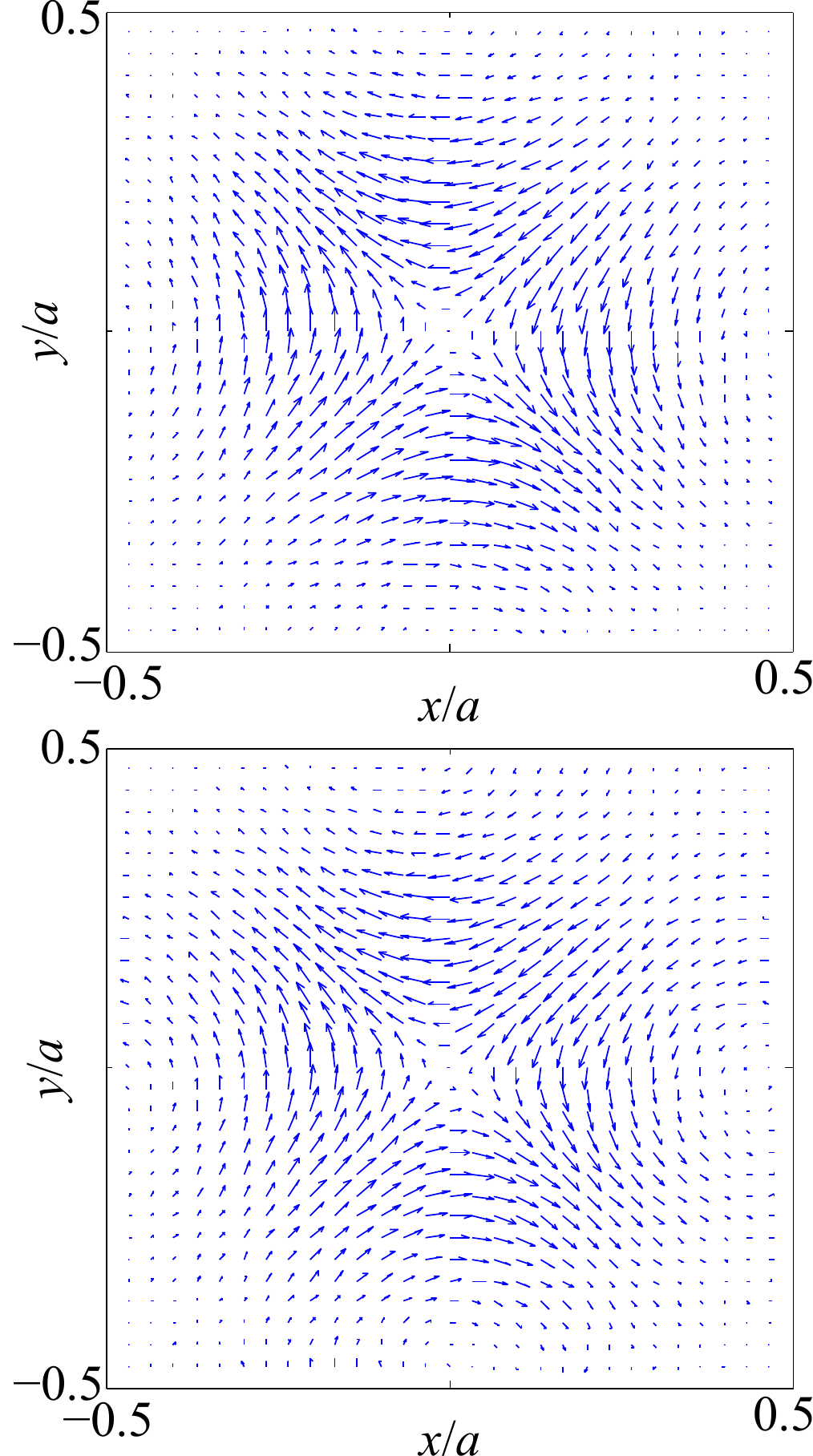} 
\par\end{centering}

\caption{\label{fig:SpinTexture} The spin textures {[}Eqs.~\eqref{eqs:spinTexture}{]}
for the Mott state (upper panel) and for the superfluid state (lower
panel) in one unit cell at the same parameters as in Fig.~2 in the
main text.}
\end{figure}

We define the spin texture as the projection of the spins onto the
$x$-$y$ plane, with the components:\begin{subequations}\label{eqs:spinTexture}
\begin{align}
S_{x}\left(\bm{r}\right) & =2{\rm Re}\left\langle b_{\uparrow}^{\dagger}\left(\bm{r}\right)b_{\downarrow}\left(\bm{r}\right)\right\rangle \\
S_{y}\left(\bm{r}\right) & =2{\rm Im}\left\langle b_{\uparrow}^{\dagger}\left(\bm{r}\right)b_{\downarrow}\left(\bm{r}\right)\right\rangle 
\end{align}
 \end{subequations} In Fig.~\eqref{fig:SpinTexture}, we plot the
spin textures in the Mott and the superfluid states.

 \bibliographystyle{apsrev4-1}
\bibliography{jabref_database_8_28_15}


\end{document}